\documentclass[twocolumn, 11pt]{article}
\usepackage[a4paper, total={7in, 10in}]{geometry}

\usepackage{abstract}
\usepackage{amssymb}
\usepackage{bm}
\usepackage{dcolumn}
\usepackage{graphicx}
\usepackage[symbol]{footmisc}
\usepackage{hyperref}
\usepackage{multicol}
\usepackage{multirow}
\usepackage{subfiles}
\usepackage{tabularx}
\usepackage{verbatim}

\usepackage{titlesec}

\titleformat{\section}
  {\normalfont\fontsize{13}{15}\bfseries}{\thesection}{1em}{}

\def\half{{\textstyle \frac{1}{2}}}

\newcommand\kappabar{\overline \kappa}
\renewcommand\k{\kappa}
\newcommand\kb{\kappabar}
\newcommand\der {\mathrm{d}}

\newcommand\e{\epsilon}

\newcommand\ak{\alpha}

\begin{document}

\title{\Large Snapping and Switching of Elastic Arches with Patterned Preferred Curvature}
\author{Michał Zmyślony, Ammar Khan\thanks{Also at Department of Physics, SBASSE, Lahore University of Management Sciences, Lahore, Pakistan}, John S. Biggins\thanks{\href{mailto:jsb56@cam.ac.uk}{jsb56@cam.ac.uk}} \\
 \small\textit{Department of Engineering, University of Cambridge,} \\
\small \textit{Cambridge, United Kingdom}
}
\date{\today}

\twocolumn[
\maketitle 
\vspace{-2.5em}
    \begin{onecolabstract}
    \vspace{-1em}
       An elastic arch is an archetypal bistable system. Here, we combine elastica theory and photo-mechanical experiments to elucidate the mechanics of an active arch with a spatio-temporally varying preferred curvature $\kb(s)$. Our shallow-arch theory completely describes any such system via the decomposition of its $\kb(s)$ into Euler-buckling modes. Intuitively, if $\kb(s)$ overlaps with the fundamental mode, it snaps the arch up/down. Conversely, non-overlapping $\kb(s)$ drives a second-order transition to a higher-order shape. Furthermore, the form of $\kb(s)$ enables control over the instability's character; we find the forms for snapping with maximum energy release and at the lowest stimulation (both binary patterns) and design forms for symmetric and asymmetric switching pathways. Analogous control can also be achieved in boundary-driven instabilities of passive arches by fabricating them with suitable $\kb(s)$. We thus anticipate our results will improve switchable/snapping elements in MEMS, robotics, and mechanical meta-materials. 
    \end{onecolabstract}
]
\saythanks

The instabilities of elastic rods have long enchanted researchers \cite{Euler_Buckling, born1906untersuchungen}, and unify slender bodies from nanotubes \cite{yakobson1996nanomechanics, pantano2004mechanics} to cross-continental pipelines \cite{hobbs1984service} and undersea cables \cite{zajac1962stability}. Celebrated examples include Euler's buckling of a compressed column \cite{Euler_Buckling},  the snap of Timoshenko's bi-metal thermostat \cite{Timoshenko_1925}, and the coiling transitions found in tendrils \cite{gerbode2012cucumber}, DNA \cite{le1984twist}, and twist-based artificial muscles \cite{haines2016new, charles2019topology}. Each such instability nucleates from a simple base-state with uniform compression/twist/natural curvature, making the complex forms that emerge a compelling morphogenetic mechanism \cite{savin2011growth, nelson2016buckling}. However, in both biology and the emerging field of shape-responsive solids, there is no fundamental restriction to translationally invariant beams: rather, patterns of growth or actuation can sculpt almost arbitrary spatio-temporal patterns of natural curvature \cite{kohn2018bending, kaczmarski2022active}. Thus, it is natural to ask whether such patterns can drive, direct, and optimize instabilities.

Here, we address this question for a beam that is pre-buckled into an elastic arch. Such arches are elementary bi-stable systems leading to applications as MEMS switches \cite{joshitha2017fabrication}, binary-sensors \cite{Timoshenko_1925}, and mechanical memories  \cite{mei2023memory}. Bi-stable switching also involves energy storage and sudden release (snapping), enabling explosive jumpers \cite{wang2023insect}, catchers \cite{Forterre_2005, Smith_2011}, and throwers \cite{Polat_2024}, and powering waves through damped mechanical meta-materials \cite{raney2016stable}. Consequently, recent studies have examined many external routes to induce snapping, including boundary manipulation \cite{Gomez_2017, Sano_2018, Wan_2020, Radisson_2023, Wang_2023}, electro/magnetic fields \cite{Yin_2007, Nadkarni_2016}, and even surface tension \cite{Fargette_2014}. 

Alternatively, arch-snaps can be driven internally by changing preferred curvature. Such snapping was first observed in bimetallic strips that bend on heating \cite{Timoshenko_1925}, and, more recently, using other strips that bend on electrical \cite{Aimmanee_2018, Rossiter_2006}, thermal \cite{Polat_2024}, or optical  \cite{Shankar_2013, Waters_2024} stimulation. In such systems, inducing uniform preferred curvature suffices to snap a simply supported (pinned) arch back and forth, as the up/down states have pure negative/positive curvature, respectively \cite{Timoshenko_1925}. Flat-clamped arches are significantly easier to fabricate, but bear opposite curvatures in their middle and sides, preventing snapping under uniform curvature stimulation. However, such arches may be snapped by stimulating the central segment \cite{Aimmanee_2018, Rossiter_2006, Polat_2024, Shankar_2013, Waters_2024}, and snapped back by reversing the sign \cite{Aimmanee_2018, Rossiter_2006, Shankar_2013} or pattern \cite{Waters_2024}. These results highlight the subtlety and importance of patterned stimulation and motivate our analysis.

\section*{Experimental system}
\begin{figure}
    \centering
    \includegraphics[width=\linewidth]{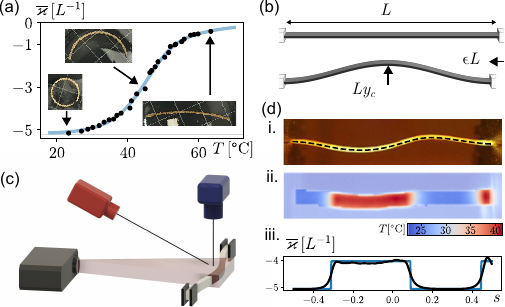}
    \caption{Experimental system. (a) Measured spontaneous curvature $\overline \varkappa $ of the active beam. (b) A beam of length $L$ is clamped and compressed by $\e L$ into an arch, and (c) mounted in front of a projector (grey), thermal (red) and visible light (blue) cameras. (d) The projector and i.\ visual and ii.\ thermal feeds form a feedback loop controlling iii.\ the curvature (temperature) profile (black: measured; blue: target). Brighter pixels in ii.\ denote temperature-based background excluded from iii.\ $\overline{\varkappa}$ extraction.
    }
    \label{fig: experimetnal info}
\end{figure}

To do so, we prepare a thermally-bending nematic elastomer strip using standard mechanical programming techniques \cite{Barnes_2019, bauman2022actuation} ($L = 40\, \mathrm{mm}$, $w=3.5\, \mathrm{mm}$, $t=0.6\,\mathrm{mm}$, App.\ \ref{app: sample preparation}). The resultant strip has a uniform natural curvature $\overline \varkappa_{max}=-5.2\,L^{-1}$ in the nematic phase (at room temperature), falling to approximately zero in the isotropic phase ($T>65 ^\circ \mathrm C$, Fig.\ \ref{fig: experimetnal info} (a)). While we choose to use nematic elastomers, similar thermal benders could be fabricated using many other systems, yielding similar results.

The strip is flat-clamped, buckled into an arch by compressing by $\epsilon=1.2\%$ (Fig.\ \ref{fig: experimetnal info} (b)), and mounted in front of a DMD-based projector (DLP 4750LC, EKB Technologies), oriented such that the natural curvature is negative (downward), Fig.\ \ref{fig: experimetnal info} (c). The projector creates a 4 cm $\times$ 2.25 cm 1080p image that entirely covers the strip, enabling arbitrary illumination patterns with an intensity of up to 750 mW/cm$^2$. The shape and temperature of the strip are monitored by visible light and thermal cameras (Fig.~\ref{fig: experimetnal info} (d)), and combined in real time (9 Hz) to reconstruct an arc-length parametrized temperature profile $T(S)$ along the strip, and therefore $\overline \varkappa(S)$ via the calibration curve, Fig.~\ref{fig: experimetnal info} (a).
 
During a particular experiment, we wish to prescribe a specific preferred curvature profile $\overline \varkappa(S, t)$. To do so, we simply compare the current target against the observed $\overline \varkappa(S)$ and apply bang-bang control, illuminating only the regions where the temperature is below target. Crucially, our real-time shape-tracking system enables us to robustly impose the preferred curvature profile in the Lagrangian/material arc-length description, despite the projector naturally operating in the Eulerian/laboratory frame, Fig.~\ref{fig: experimetnal info} (d). 

\section*{Snapping of centrally stimulated arch}
\begin{figure}[t]
    \centering
    \includegraphics[width=\linewidth]{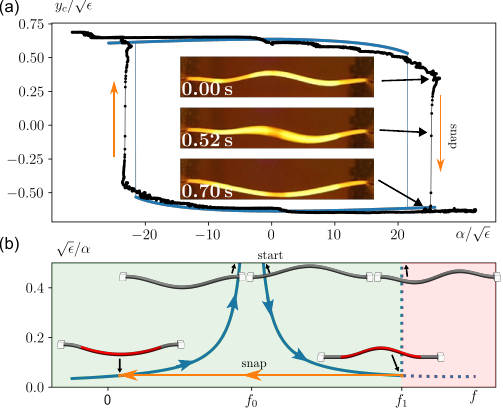}
    \caption{Snapping of arches with active curvature. 
    (a) Theoretical (blue) and experimental (black) hysteresis loop of center height vs curvature magnitude showing both way snapping control under the stimulation of central $70\%$. (b) Theoretical plot of normalized arch compression  $\sqrt \e / \alpha$ for each arch state as a function of longitudinal force $f$, showing compression peaks at the Euler buckling forces $f_i$, and a transition from stable (solid) to unstable (dashed) solutions at $f_1$, corresponding to the snap observed at fixed $\epsilon$ and increasing stimulation.}
    \label{fig1}
\end{figure}

We begin by stimulating the central $70\%$ of the strip to induce an approximately piecewise-constant curvature profile with background $\overline \varkappa_0$ in the outer pieces and an increasing $\overline\varkappa=\overline \varkappa_0+\alpha L^{-1}$ in the center. The beam snaps down at amplitude $\alpha=2.75$ and back up at $\ak=-2.56$ (achieved by stimulating the edges), leading to the hysteresis loop in Fig.\ \ref{fig1} (a) and Mov.\ 1. The experimental data were extracted from a 50 FPS recording during post-processing, enabling us to observe the antisymmetric character of the snapping transition.  

To analyze these snaps, we model the arch as an inextensible elastica with bending stiffness $B$, length $L$ and (arc-length parametrized) preferred curvature profile $\overline\varkappa(S)$, such that, if it follows the plane curve with tangent angle $\theta(S)$ and local curvature  $\varkappa(S)=\theta'(S)$, it stores bending energy 
\begin{equation}
E = \frac{1}{2}B\int_L (\varkappa-\overline\varkappa)^2\, \der S.
\end{equation}
We note this 1D energy naturally emerges for our narrow strips (transversely relaxed) or very wide sheets (transversely suppressed), but intermediate widths require a full 2D plate treatment \cite{Polat_2024}. Morphology and stability then follow from energy minimization. Unconstrained, the elastica can trivially follow $\overline \varkappa$ to achieve $E=0$, but the flat-clamps impose the end-to-end displacement, $\langle \sin{\theta}\rangle\equiv \frac{1}{L}\int_L \sin{\theta} \,\mathrm{d}S=0$ (vertical) and $\langle 1-\cos{\theta}\rangle=\epsilon $ (horizontal). Minimizing variationally subject to these constraints yields the non-linear ODE $B \,\theta'' + F \sin \theta+V \cos \theta= B\, \overline\varkappa'$, where the horizontal and vertical clamp forces $F$ and $V$ arise as Lagrange multipliers. Flat clamping further imposes $\theta=0$ boundary conditions, while freely-rotating (pinned) ends require $\theta'=\overline\varkappa$. Hence, pinned arches respond to uniform $\overline\varkappa$, while clamped arches require patterns with $\overline\varkappa'$.

To reproduce the experiment, we consider a flat-clamped arch compressed by $\epsilon=1.2\%$ and set  $\overline\varkappa(S)$ to a binary pattern, with a uniform but growing value in the central 70\% and zero otherwise. The system's nonlinearity precludes an analytic solution, but the resulting configurations and their linear stability can be readily solved using SciPy's solve\_bvp (Sec.\ S1.1). Alternatively, arch configurations and (damped inertial) dynamics can also be simulated using a discrete rod model (Sec.\ S1.2). Both fully non-linear approaches confirm two-way snapping via an asymmetric mode at $\ak=\pm2.36$; equivalent snap-back can also be achieved using the dual stimulation, shown in Mov.\ 1, as this also exactly reverses $\overline\varkappa'$.

To gain analytical insight we non-dimensionalize  ($S=Ls, \varkappa=\kappa/L,F=B\,f /L^2,V=B\,v/L^2, E=B\mathcal{E}/L$), and expand to leading order in $\theta$ to obtain the light-compression equations:
\begin{equation}
 \theta'' + f \theta +v =  \kb',\mathrm{\ \ } \theta(\pm \half)=0,\mathrm{\ \ }\langle\theta\rangle=0,\mathrm{\ \ } \langle \half \theta^2\rangle= \epsilon. \label{eq:EL_shallow}
\end{equation}
During the pre-buckling, we have $\kb'=0$. The first three equations then form Euler's classic buckling eigen-system, with modes at discrete compressive forces $f_i$  and corresponding orthonormal mode shapes, $\langle \half \theta_i \theta_j\rangle=\delta_{ij}$ with wavenumber $k_i=\sqrt{f_i}$ and alternating even ($\sin(k_i/2)=0$) and odd ($\tan{(k_i/2)}=k_i/2$) morphologies. Finally, the non-linear compression constraint fixes amplitude, $\theta=\pm \sqrt{\epsilon} \theta_i$, with two mirrored solutions at each $f_i$. The higher order modes all require $f>f_0$, and are thus unstable to growth of $\theta_0$ \cite{Nayfeh_2008}, so the pre-buckled arch must sit bi-stably in one of the two $i=0$ modes.

\begin{figure*} \label{fig2}
    \centering
    \includegraphics[width=\linewidth]{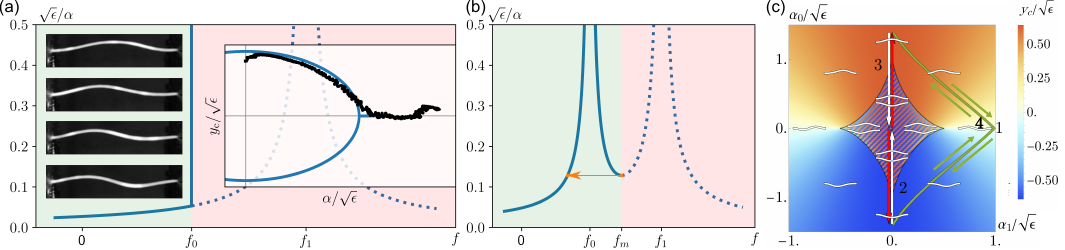}
    \caption{Snapping condition and subcritical deformation. (a) Compression spectrum and (inset) center heights of an arch stimulated with $\kb_1$. (b) Snapping at $f_m<f_1$ by stimulating with $\kb= \alpha_0 \k_0+\alpha_1 \k_1$ and (c) the corresponding phase diagram showing bistability in the dashed region and monostability outside.    }
\end{figure*}

We next consider introducing a top-hat preferred curvature $\kb_\mathrm{rect}=\ak\, \mathrm{rect}(s/\gamma)$ that stimulates the central fraction $\gamma$ of the arch and with amplitude $\ak$. Such stimulation introduces a source term into the shape-equation, and the linear system now admits an analytic particular integral $\theta_\mathrm{rect}(s,f)\propto \ak $ for every compressive force $f$ (Sec.\ S2.8).  Furthermore, if $f$ coincides with one of the original buckling forces, $f_j$, we must add the associated buckling form, $ c\, \theta_j$, as a complementary function. Inserting these expressions into the compression equation, we obtain an algebraic force-amplitude relation $\sqrt \epsilon/\ak= g(f)$, shown in Fig.\ \ref{fig1} (b) for $\gamma=0.7$. This ``spectrum'' maps the compression generated by each solution and has a continuous background from $\theta_\mathrm{rect}$, augmented by degenerate spikes at the $f_i$ parametrized by the constant of integration $c$. At a given compression and stimulation $\sqrt \epsilon/\ak$ is fixed and corresponds to a horizontal line on the Fig.\ \ref{fig1} (b) plot, which starts from $\sqrt \epsilon/\ak\to \infty$ and lowers as stimulation increases (Mov.\ 1). Its intersections with $g(f)$ are then the solutions describing precise arch morphologies available at that $\sqrt \epsilon/\alpha$.

At the start of the experiment, the high horizontal line intersects only in pure buckling modes, and the arch is pre-buckled in the fundamental mode ($i=0$). As stimulation increases, the line descends and the solution ``slides'' down the left or right tail of the $f_0$ divergence, depending on whether the initial state accords (left) or opposes (right) the new preferred curvature. According preferred curvature reduces the required compressive force, ultimately placing the arch in tension. Conversely, opposing $\kb$ increases the compressive force,  until the solution eventually reaches the (unsplit) spike at $f_1$ ($\ak=21.52\sqrt{\epsilon})$, where it encounters a bifurcation and the arch becomes linearly unstable to growth of $\theta_1(s)$, and snaps to the stable ``down'' configuration at the same compression. This simple analysis thus predicts arch morphology, snapping threshold and snapping mode. We find that all these predictions agree well with our experiment, and, at $\epsilon=1.2\%$, essentially perfectly with our fully non-linear discrete-rod numerics and with only minor disagreements at higher compressions (Fig.\ S4). We note that our experimental system experiences minor length changes during stimulation due to changes in compressive force, App.\ \ref{app: compression}, which can be extracted from arch profiles and are reported as minor $\e$-variations in Supplementary Movies. However, the arch is sufficiently slender that our inextensible model nevertheless provides excellent agreement.

\section*{Response to general stimulation}
More generally, any preferred curvature profile can be decomposed as a sum of buckling curvatures, $\kappa_i\equiv \theta_i'$, plus an irrelevant affine component (Sec.\ S2.3). An individual component $\alpha_i \kappa_i$ then contributes a particular integral for the linear equations $\theta_\mathrm{PI}(s,f)=\alpha_i \theta_i(s) f_i/(f-f_i)$ and accommodates compression (Sec.\ S2.1)
 \begin{equation}
\epsilon= \alpha_i^2 \frac{f_i^2}{(f-f_i)^2}.\label{eq:compression_component}
\end{equation}
Mode orthogonality allows these compression contributions to be simply summed to obtain a total continuous compression spectrum with a divergence at each included $f_i$, augmented by degenerate spikes at any excluded $f_i$ via the complementary functions. Thus, the even central top-hat stimulation has continuous divergences at each even buckling mode and spikes at odd ones, including $f_1$ where snapping commences, Fig.\ \ref{fig1} (b). 

Stimulating just the fundamental mode, $\kb=\alpha_0 \k_0$, leads to a simple spectrum with only the $f_0$ spike split. Snapping proceeds similarly at $f_1$, though now at threshold stimulation $\alpha_0 = 1.05 \sqrt \epsilon$ ($\ak=26.28\sqrt{\epsilon}$). Conversely, stimulation profiles orthogonal to $\k_0$ (e.g., $\kb=\k_1$) yield a very different spectrum, with a degenerate spike at $f_0$, as stimulation no longer distinguishes the fundamental up/down states. Such an experiment still starts pre-buckled into the fundamental mode, so on stimulation the solution ``slides'' vertically down the $f_0$ spike, with a corresponding reduction in its  $\theta_0$ component, until reaching the bottom of the spike at $\alpha_1 = 0.51$ ($\ak = 18.84\sqrt \e$) and turning left into the $f<f_0$ continuous solutions that are pure particular integral, and hence orthogonal to $\theta_0$. Thus, under increasing stimulation, we expect a continuous (second-order) morphological transition from solutions that overlap with $\theta_0$ to those that do not. Such continuous transitions describe how complex $\kb$ can morph an arch into normally unstable higher-order morphologies. 

To observe such a transition, we apply a $\kb=\alpha_1 \k_1$ profile to the arch and track the displacement of the central point ($s=0$) above the clamps. Its position provides a convenient measure of overlap between the arch shape and the fundamental mode, as every higher-order mode has $y_c=0$. As expected, both the dynamic numerical simulation and experiment confirm that this displacement vanishes continuously above a threshold stimulation in precise agreement with theory, and without any sudden snap, inset Fig.\ \ref{fig2} (a) and Mov.\ 2.

Having understood stimulation in pure zeroth and first modes, we now analyze their combination $\kb=\alpha_0 \k_0 + \alpha_1 \k_1$, which leads to divergences at $f_0$ and $f_1$ as shown in Fig.\ \ref{fig2} (b). In this case, increasing stimulation amplitude at fixed $\alpha_0/\alpha_1$, the solution slides down the right side of the $f_0$ peak, until we eventually encounter a minimum between the two divergences, $f_0<f_m<f_1$, at which point, no further ``up" solutions are available, and the arch must snap down. Adding  $\k_1$ thus expedites snapping, as it directly stimulates to the previous snapping pathway.    

We validate this graphical intuition by considering the stability of a base solution  $\mathbf{w}_0(f)=(f,v(f),\theta(s,f))$ to small perturbations $\delta \mathbf{w}_1$. To obtain the equations of linear stability, we first substitute $\mathbf{w}_0+\delta \mathbf{w}_1$ into Eq.\ \ref{eq:EL_shallow} and linearize in $\delta$. Comparing the linear terms with the derivative of the zeroth-order terms with respect to $f$ confirms linear instability at $f_m$, to perturbations of the form $\mathbf{w}_1=\partial_f \mathbf{w}_0(f)$ (Sec.\ S2.4). 

These results may be summarized on a phase diagram of stable states in $(\alpha_0/\sqrt{\epsilon}, \alpha_1/\sqrt{\epsilon})$ space, as shown in Fig.\ \ref{fig2} (c). The diagram naturally contains a large bistable region around the origin (high compression, low stimulation), corresponding to up and down arches with a large component of $\theta_0$. These snap to mono-stable at sufficient $\alpha_0$, while introducing $\alpha_1$ reduces the threshold and amplitude of the snaps, until they disappear entirely via mechanical critical points at $\left(0,\pm (1-f_0/f_1)^2\right)$. This structure highlights an additional possibility: by choosing a suitable route through the diagram, one can move from the up-state at (0,0) to the corresponding down state by traveling around the critical point, and hence without any discontinuous snap. An experimental implementation of this possibility is shown in the Mov.\ 3 and described in detail in Sec.\ S2.9, with the followed $(\alpha_0, \alpha_1)$ route marked in Fig.\ \ref{fig2} (c).

\section*{Optimized snapping}
\begin{figure}
    \centering
    \includegraphics[width=\linewidth]{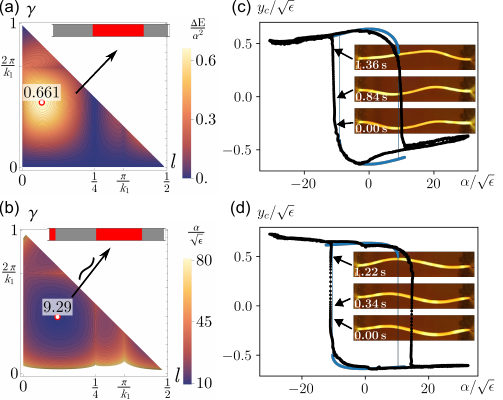}
    \caption{Optimized stimulus profiles. (a) Elastic energy released during the snap and (b) snap-through thresholds for offset rectangular $\kb$, inset:  optimal profiles. Comparison of center heights for (c) energy optimized $\kb$ and (d) threshold optimized $\kb$, theoretical design (blue) and experiment (black).}
    \label{fig3}
\end{figure}
Having understood the fundamentals, we now search for preferred curvature profiles that optimize snapping and switching. We first consider an arch that can attain the maximum curvature $0<\kb(s)<\ak$, and ask what pattern and $\epsilon$ should be chosen to maximize energy release, as desired in explosive applications such as jumping. To do so, we start numerically by writing the preferred curvature as a (finite) sum of buckling modes, then using SciPy to rapidly analyze the associated snap (via Eq.\ \ref{eq:compression_component}) and optimize $\Delta \mathcal{E}$ over mode amplitudes.  Optimization uncovers a binary pattern, with $\ak$ stimulation in an offset central segment, $\kb =\ak \,\mathrm{rect}((s-l)/\gamma)$. Inspired by this result, we solve the problem analytically for this class of profiles and optimize over stimulation width $\gamma$ and its offset $l$. Both approaches find the same optimum which releases $\Delta \mathcal{E} =0.661\,\ak^2$ at $\ak = 10.33 \sqrt \epsilon$, compared to $\Delta \mathcal{E} =0.239\,\ak^2$ for pure $\k_0$ snapping, Fig.\ \ref{fig3} (a). 

Conversely, in switching applications, we are interested in the profile that switches at a minimum $\ak$. In the design space of offset rectangular stimuli we find the minimum threshold of $\ak = 9.29\sqrt \epsilon$, substantially below $\ak=26.28 \sqrt{\epsilon}$ for pure $\k_0$ stimulation, Fig.\ \ref{fig3} (b). Repeating the general numerical optimization reveals the threshold may be lowered slightly further  ($\ak=9.11 \sqrt{\epsilon}$) by introducing an additional active segment at the arch's edge. Experimentally, we observe good agreement in the thresholds/hysteresis for both optimized profiles, Fig.\ \ref{fig3} (c), (d) and Mov.\ 4, 5.

Moreover, the threshold plot (Fig.\ \ref{fig3} (b)) of the offset rectangular profile reveals great sensitivity of central stimulation ($l=0$) to misalignment, with even small offsets dramatically reducing the threshold as they introduce direct coupling to the $\kappa_1$ mode. However, this sensitivity is completely suppressed if the  stimulation width matches the wavelength of the 1st mode ($\gamma=2\pi/k_1\approx0.7$), decoupling the stimulation from $\kappa_1$ and leading to an obvious plateau on the threshold plot, and enabling predictable snapping. Analogous features are observed at $l=\pi/k_1$ and $l=\pi/k_0$, associated with decoupling from $\k_1$ and $\k_0$ respectively, the latter leading to perfect supercritical behavior. Conversely, both optimal stimuli lie in smooth energy/threshold minima, making them insensitive to imperfections. 

\section*{Symmetric snap-through}
Lastly, we explore the use of preferred curvature to modify the snapping pathway. In previous examples with symmetric stimulation (central-segment, $\k_0$), snap-through occurred at $f_1$ via the 1st (antisymmetric) mode. Similarly, if the preferred curvature contains $\k_1$, even the stable stimulated arches (particular integral) have an antisymmetric $\theta_1$ component, and when snapping occurs at the spectral minimum $f_0<f_m<f_1$, the growing mode $\partial_f \theta(s,f)$ directly inherits this (lack of) symmetry. In both cases, we thus observe antisymmetric displacements and rotation of the center-point, which may be undesirable in applications where this center-point interacts with other elements. Can we instead drive a snap-through that proceeds entirely symmetrically?

\begin{figure}
    \centering 
    \includegraphics[width=\linewidth]{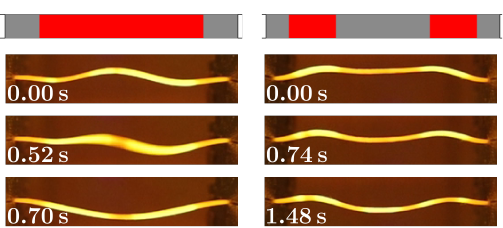}
    \caption{Controlling symmetry of snap-through transition.  Stimulation profiles $\kb_\mathrm{rect}|_{\gamma=0.7}$ and $\kb_\mathrm{sym}$, and the corresponding mid-snap images.}
    \label{fig4}
\end{figure}

To achieve such symmetric snapping, we consider a symmetric profile $\kb=\alpha_0 \k_0+ \alpha_2 \k_2+...$, that generates a symmetric particular integral with compression divergences at $f_0$ and $f_2$, and hence a minimum in between. A snap at this minimum's bifurcation will inherit the base-state's symmetry. However, it will only be observed if it precedes the standard asymmetric snap, $f_m<f_1$. Fortunately, the minimum may be moved arbitrarily close to $f_0$ by reducing $\alpha_0$, though at the cost of reducing the splitting of the $f_0$ peak and, hence, snapping strength. To demonstrate such snapping, we design a pattern composed of two segments, each spanning $20\%$ of the arch's total length and centered at $s=\pm0.3$, resulting in a symmetric snap-through, as shown in Fig.\ \ref{fig4} and Mov.\ 6. Moreover, even in the case of central stimulation from Fig.\ \ref{fig1} for which the symmetric and antisymmetric thresholds are in proximity ($\ak^\mathrm{sym}\approx1.1\ak^\mathrm{asym}$), during the edge stimulation in Mov.\ 1 we observe a symmetric snap-back, most likely due to reduction in $\alpha_0$ and increase in $\alpha_2$ caused by thermal diffusion.

\section*{Discussion}

Although our analysis focuses on actively curving arches, there is a duality between increases in preferred curvature and decreases in compression; therefore, our model readily extends to passive pre-curved arches under active boundary loading (Sec.\ S2.11). Moreover, our flat-clamped results directly generalize to any combination of flat-clamped and pinned arches when analyzed in terms of the relevant buckling modes (Sec.\ S2.6). However, much of the previous literature focuses on snapping of $\kb=0$ arches with ends clamped at finite angles $\theta_L$ and $\theta_R$ and snaps induced by changing $\epsilon$. To connect with these results, we note that such clamping conditions themselves generate a particular integral equivalent to a preferred curvature with mode amplitudes $\alpha_i/4=\theta_L+\theta_R (i \mathrm{\ even}); \theta_L-\theta_R (i \mathrm{\ odd})$, and we may use the corresponding compression spectrum to analyze snapping. In particular, this analysis immediately explains why asymmetric clamping control results in continuous transitions, while symmetric clamping enables snap-through following an antisymmetric pathway \cite{Wan_2020, Radisson_2023}. Furthermore, in the latter case, the symmetric transition mode lies just beyond $f_1$, enabling its observation dynamically \cite{Wang_2023} if not quasistatically. However, clamping angles can provide only 2-DOF control over snapping, thereby vastly reducing the potential for optimization.

Overall, our approach thus provides a comprehensive and straightforward understanding of snapping in arches, illuminates prior results on naturally straight passive arches, and elucidates the more complex behavior of arches with preferred curvature, whether fixed at fabrication or induced by stimulation. Fabricated curvature enables tuning of thresholds, energy release, and mode shape, paving the way for improved snapping in metamaterials and other externally snapped devices. Arches with dynamic preferred curvature drive their own snaps, and can additionally sustain unexpected sequences of transitions and instabilities, including morphing between higher-order forms, non-reciprocal motion, and switching without snapping at all. Our experimental realizations of these phenomena are all demonstrated in a single arch, hinting at how such control may enable complex robotic functions and highlighting how mechanical constraints enrich and complicate active morphing.

Our work also inverts a common paradigm: we do not form patterns from instabilities, but rather instabilities from patterns. Looking ahead, this inversion has broad applicability beyond snapping arches, leading to similar opportunities for optimization and enrichment. Encouragingly, in biology, it is increasingly understood that evolution often harnesses pre-patterning to regulate and guide buckling morphogenesis \cite{bard1982morphogenesis, tallinen2016growth, shyer2015bending}; similarly, the celebrated snapping in Venus flytraps \cite{Forterre_2005} and hummingbird beaks \cite{Smith_2011} will inevitably have already undergone evolutionary optimization over stimulation patterns and initial shape. We thus anticipate a rich seam of research on the interaction of patterns and instabilities, and hope the elementary system analyzed here provides a useful foundation.

\section*{Acknowledgments}
This work is funded by the European Union’s Horizon 2020 Research and Innovation Programme under the Marie Skłodowska-Curie Grant Agreement No. 956150 (STORM-BOTS). Additionally, J.S.B. and A.K. received funding from a UKRI ‘Future Leaders Fellowship’ grant (grant nos. MR/S017186/1 and MR/Y033957/1).

M.Z. conducted the theoretical and numerical investigation; A.K. conducted the experiments; M.Z. and J.S.B conceptualized the work. All authors contributed to writing the manuscript.

All collected data and software required for reproduction of the results of this publication are publicly available on Zenodo \cite{zenodo_dataset}.

\appendix
\section{Sample preparation} \label{app: sample preparation}

Mesogenic diacrylates RM257 (CAS: 174063-87-7) and C6BAPE (CAS: 151464-39-0) were acquired from Daken Chemical.  Dithiol chain extender EDDET (CAS: 14970-87-7),  tetra-thiol crosslinker PETMP (CAS: 7575-23-7), catalyst dipropylamine DPA (CAS: 142-84-7),  photoinitiator Irgacure 2959 (CAS: 106797-53-9), absorbing dye Brilliant Yellow (CAS: 3051-11-4), and solvent chloroform (CAS: 67-66-3) were acquired from Sigma Aldrich. All reagents were used as received. 

We fabricated LCE samples via mechanical programming, following a protocol analogous to \cite{Barnes_2019, bauman2022actuation}, with molar ratios of RM257, C6BAPE, EDDET and PETMP of 0.55-0.55-0.75-0.125, to maintain an excess of acrylate bonds. Photoinitiator was added in 1.5 wt\% together with 0.25\% photo-dye and 20 wt\% solvent to form an isotropic solution. First-stage polymerization was initiated by adding a 10 wt\% catalyst solution (5 vol\% DPA in chloroform) in a 700 \textmu m-thick glass cell and then left to complete at room temperature for 18 h, fixing the flat geometry as the isotropic configuration. The sample was taken out of the cell, the solvent evaporated at $100^\circ\mathrm{C}$ for 2 h, yielding a 600 \textmu m-thick sheet, which was cut into a strip measuring $48\,\mathrm{mm}\times3.5\,\mathrm{mm}$, with an excess length for clamping. The strip was then mechanically programmed by wrapping it around a cylindrical mold of radius $R_\mathrm{prog}=7.6\,\mathrm{mm}$ fixed in via exposure to UV light, imprinting uniform natural curvature $\varkappa_\mathrm{prog}= 1.3\,\mathrm{cm}^{-1}$ at room temperature. Upon heating, the material undergoes nematic-to-isotropic phase change and nearly recovers the flat isotropic geometry at $T>65^\circ\mathrm {C} $.

\section{Observed compression variation} \label{app: compression}
Experimental data analysis detects a small variation in $\epsilon$ throughout the experiment due to the extensibility of the active beam. The length change due to compressive force is given by 
\begin{equation} \label{eq: compression change}
    \Delta \epsilon =-\frac{Y t^3 \, w / 12}{Y t\, w \, L} \frac{f}{L} = -\frac{t^2 f}{12 L^2},
\end{equation}
where $Y$ is the Young's modulus, $t$ the beam thickness, $L$ its length and $f$ the non-dimensionalized compressive force. 

For the dimensions of our beam ($t = 0.6$ mm, $L = 40$ mm) and prior to any stimulation, i.e. $f = f_0$, we calculate $\Delta \epsilon =-0.075\%$ and extract from Mov.\ 1 $\epsilon=1.2\%$. At the snap-through, we have $f=f_1$ giving $\Delta \epsilon =-0.15\%$ compared to the extracted $\epsilon=1.14\%$, while after the snap $f\approx0$ gives $\Delta \epsilon =0\%$ and $\epsilon=1.28\%$, giving satisfactory agreement between the theory and experiments. 

\bibliographystyle{abbrv}
\bibliography{references}% Produces the bibliography via BibTeX.

% \subfile{supplementary_information}
\end{document}

\maketitle
\tableofcontents
\newpage

\section{Solving the non-linear problem}
In order to obtain the Lagrangian, we start from the 1D bending energy
\begin{equation}
    E = \frac{1}{2}B\int_L(\theta'(S)-\overline\varkappa(S))^2 \der S,
\end{equation}
to which we add the compression constraint $\int_L\cos\theta\,\der S=L(1-\epsilon)$ with the Lagrange multiplier $F$, and zero vertical displacement $\int_L\sin\theta\,\der S$ with multiplier $V$. This results in the Lagrangian being 
\begin{equation} \label{eq: SI EL nonlinear}
    \mathcal L=\frac{1}{2}B (\theta'-\overline\varkappa)^2+F(\cos\theta - (1-\e))+V\sin \theta,
\end{equation}
with the corresponding E-L equation
\begin{equation}
    B(\theta''-\overline \varkappa')+F\sin\theta+V\cos\theta =0.
\end{equation}

In addition to the two end-displacement constraints, there are boundary conditions, with three particular cases --- flatly clamped ($\theta(\pm L/2)=0$), pinned ($\theta'(\pm L/2)=\overline\kappa(\pm L/2)$, and a mixture of flat clamping on one end and pinning on the other.

Importantly, the preferred curvature enters E-L only through its derivative $\overline \varkappa'$, therefore, for clamped BCs where the BCs do not impose boundary curvature but boundary angle, any constant offsets in curvature have no effect on the deformation.

\subsection{Solving the non-linear boundary value problem using SciPy} \label{sec: SI non-linear}
The Euler-Lagrange equation, together with the boundary conditions, constitutes a nonlinear problem that cannot be solved analytically, and numerical methods must be used. In order to identify the threshold for snap-through, we first identify equilibrium configurations $\theta(F)$ at fixed $\e$, and various stimulus peak-to-peak amplitudes $\ak$ using the solve\_bvp routine from SciPy \cite{Scipy_2020}. This routine requires the problem to be posed as a first-order ODE system, which we pose using $ (x, y,\theta, \kappa)$ with two parameters $f$ and $v$ such that

\begin{gather}
    (x, y,\theta, \kappa)' = (
    \cos \vec \theta, \sin \vec \theta, \vec \k, -v \cos \theta -f\sin\theta + \ak\,\kb'), 
\end{gather}
with the following boundary conditions: 
\begin{gather}
    x(-1/2) = 0, \quad x(1/2)=1-\e, \quad
    y(\pm1/2) = 0, \quad \theta(\pm1/2)=0.
\end{gather}

These equations are solved for progressively increasing $\ak$, giving us $\vec v(\ak)=(\theta,f,v)(\ak)$ at discrete values of $\ak$, which we then interpolate in between to give a continuous function $\vec v (\ak)$.

Next, we conduct linear stability analysis by analyzing a perturbation to the equilibrium of $(x_p, y_p,\theta_p,\kappa_p)$
\begin{gather}
    (x_p, y_p,\theta_p,\kappa_p)' = (-\theta_p\sin \theta, \theta_p\cos \theta,  \k_p, \sin\theta(v \, \theta_p - f_p)- \cos\theta(f \, \theta_p + v_p)), 
\end{gather}
with BCs
\begin{gather}
    x_p(\pm1/2) =  y_p(\pm1/2) = \theta(\pm1/2)=0, \quad \k_p(-1/2)=1,
\end{gather}
where the last boundary condition is there to prevent trivial solutions. 

The solution to this problem identifies the value of $\ak$ that allows the existence of non-trivial $\theta_p$, which can then grow in amplitude to drive the instability, and determines the threshold. To improve accuracy, the process of finding discrete equilibria $\vec v(\ak)$, interpolation, and stability analysis is repeated in the vicinity of identified instability thresholds.

\begin{figure}
    \centering
    \includegraphics[width=\linewidth]{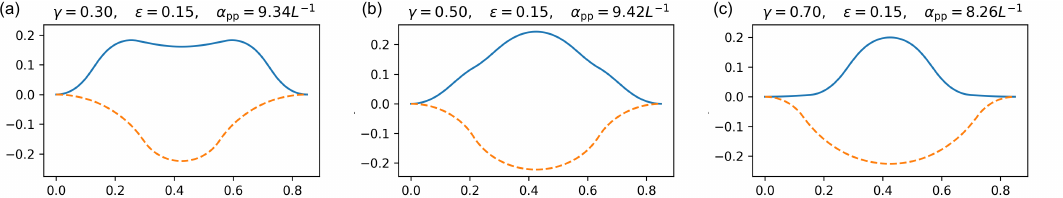}
    \caption{Beam profiles prior (solid blue) and past (dashed orange) the snap-through for high compressions obtained through non-linear optimization using solve\_bvp routine.}
    \label{fig: SI - high compression profiles}
\end{figure}

\subsection{Discrete rod model} \label{sec: SI 1D dynamics}
Alternatively, the same problem may be solved using a dynamics model. To achieve that, we modeled the beam to be composed of 101 free nodes (100 free segments) with one more node/segment at each clamped end. The beam had a normalized length of $L=1$ and experienced both bending and stretching forces, with the thickness taken to be $t=1.5 \times10^{-4}$ to enforce inextensibility. The energy is defined as 
\begin{equation}
    E=\frac{Y}{2} t\sum_{segments} \overline l_i^s\left(1-\frac{l_i^s}{\overline l_i^s}\right)^2+\frac{Y}{2} \frac{t^3}{12}\sum_{nodes}\overline  l ^n _i (\kappa_i^n-\kb_i^n)^2,
\end{equation}
where a node curvature is calculated from the difference between the segment angles, and the forces experienced by each node are $f_i=\nabla_iE$. The simulation proceeds in small time increments, with viscous damping introduced to improve convergence and stability, and terminates once all node forces and velocities are below their respective thresholds. 

The search for instability proceeds similarly to the EL approach: the beam is stimulated with progressively stronger stimuli and allowed to equilibrate before the next increment in stimulus strength. While this approach is substantially slower than the solve\_bvp approach, it enables analysis of how the beam's profile develops during the snap-through transition, which we use to present the theoretical transition path in our Movies.

\subsection{When does the 2D problem becomes 1D} \label{sec: SI Morphoshell}
Both numerical approaches above rely on the assumption that the 1D bending energy accurately represents the system. However, we know that the strip can adopt transverse curvature, which substantially modifies the strip's behavior \cite{Polat_2024}. To identify whether the dimensions of our experimental system will justify the use of a 1D model, we conducted 2D simulations using \emph{Morphoshell} \cite{Duffy_2020}. The samples were taken to have $L=40\, \mathrm{mm}$, $w=1-40\, \mathrm{mm}$, with thickness of $t=650\, \mathrm{\mu m}$, with average element size with edge length of $l = L / 75$, stiffness of $Y = 30\, \mathrm{MPa}$ and density of $\rho=1305\, \frac{\mathrm{kg}}{\mathrm m^3}$. The buckling induced end-to-end compression of $\epsilon=2\%$.  

The sample, in addition to elastic forces, also experiences a gravitational force in the side direction (side mounting) to reflect the physical system better. We note that the validity of simulations in the lower end of our thickness range is uncertain, as \emph{MorphoShell} simulates the deformation using the mechanics of thin sheets with $t\ll L_x, L_y$ \cite{efrati2009elastic}. Yet, the use of MorphoShell enables the identification of out-of-plane snapping modes, which are outside the scope of 1D theory and could facilitate snap-through before the 1D instability occurs.

\begin{figure}
    \centering
    \includegraphics[width=4 in]{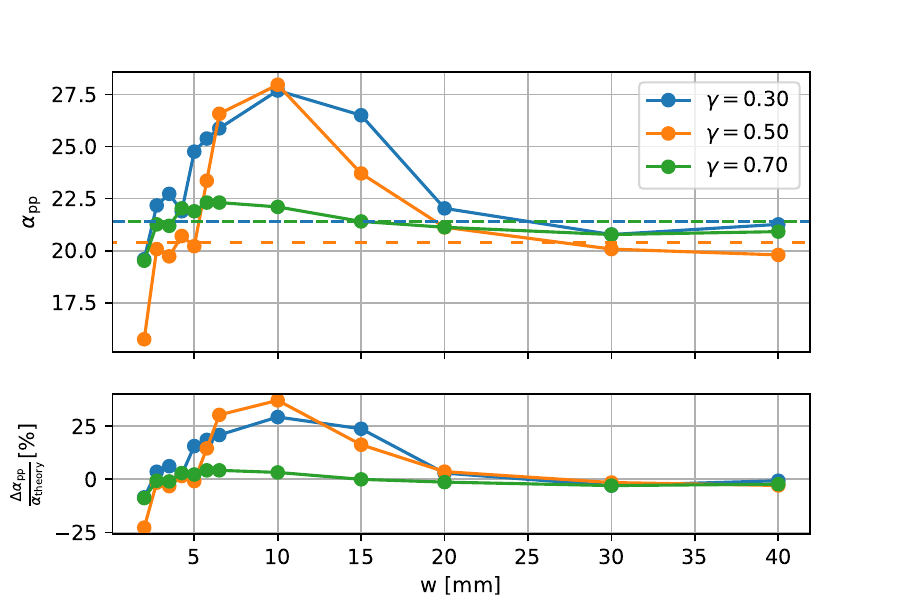}
    \caption{Comparison of the snap-through thresholds for rectangular stimuli obtained through 2D simulation using MorphoShell and 1D theory. The dashed lines represent the snapping threshold predicted by the 1D theory.}
    \label{fig: SI 2D vs 1D shallow}
\end{figure}

As shown in Fig.\ \ref{fig: SI 2D vs 1D shallow}, the 1D model accurately predicts the snapping thresholds for both narrow and wide strips. Due to constraints in our actuating system, we decided to work in the narrow regime with $t = 3.5\, \mathrm{mm}$, which lies comfortably in the region of agreement.

The snap-through threshold was identified by simulating the beam for progressively increasing values of $\ak$, equilibrating after each increment, and using the previous result as an ansatz, following essentially the same procedure as the discrete rod model. Once the snap-through was identified, we refined the search by subdividing the increment, enabling us to determine the threshold with high precision. However, a threshold found in this manner is prone to overestimation, as it depends on the equilibration threshold --- if the threshold is set too high, the runtime can increase drastically. At the same time, if set too low, the simulation will overshoot the threshold.

\clearpage
\section{Linearized shallow problem}
We begin by non-dimensionalizing the problem ($S=Ls, \varkappa=\kappa/L,F=Bf / L,V=Bv / L, E=B\mathcal{E}/L$), and expanding all the terms in E-L equation and BCs to the leading order, resulting in
\begin{align}
    B(\theta''-\overline \varkappa'+F\sin\theta+V\cos\theta =0\quad&\longrightarrow\quad \theta''+f\,\theta+v=\kb',\\
    \int_L\sin\theta\, \der S = 0 \quad &\longrightarrow \quad \int_{-1/2}^{1/2}\theta\, \der s = 0, \\
    \int_L\cos\theta\, \der S = L(1-\e) \quad &\longrightarrow \quad \frac{1}{2}\int_{-1/2}^{1/2}\theta^2\, \der s = \e,
\end{align}
while the clamping/pinning BCs remain unchanged. 

\begin{figure}[b]
    \centering
    \includegraphics[width=\linewidth]{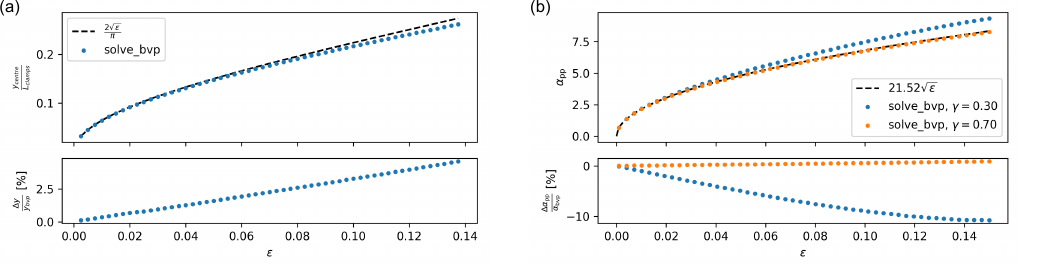}
    \caption{Comparison between non-linear solve\_bvp results and shallow angle approximation. (a) Center height in Euler buckling ($\kb=0$) and (b) snapping thresholds for a $\kappa_\mathrm{rect}$ with $\gamma=0.3$ and $\gamma=0.7$ for $\epsilon=0-0.15$ range. }
    \label{fig: SI non-linear thresholds 70} 
\end{figure}

The problem is then solved following the same procedure as in the non-linear case: first equilibria are identified, and then their stability is considered. For comparison between the approaches, Fig.\ \ref{fig: SI non-linear thresholds 70} presents the snap-through thresholds for $\kb_\mathrm{rect}$ of width $\gamma =0.3$ and $\gamma=0.7$. In the shallow limit, both stimulus widths ought to share the same threshold, yet the non-linear calculation reveals that for larger values of $\e$ they start to diverge. 

\begin{figure}
    \centering
    \includegraphics[width=\linewidth]{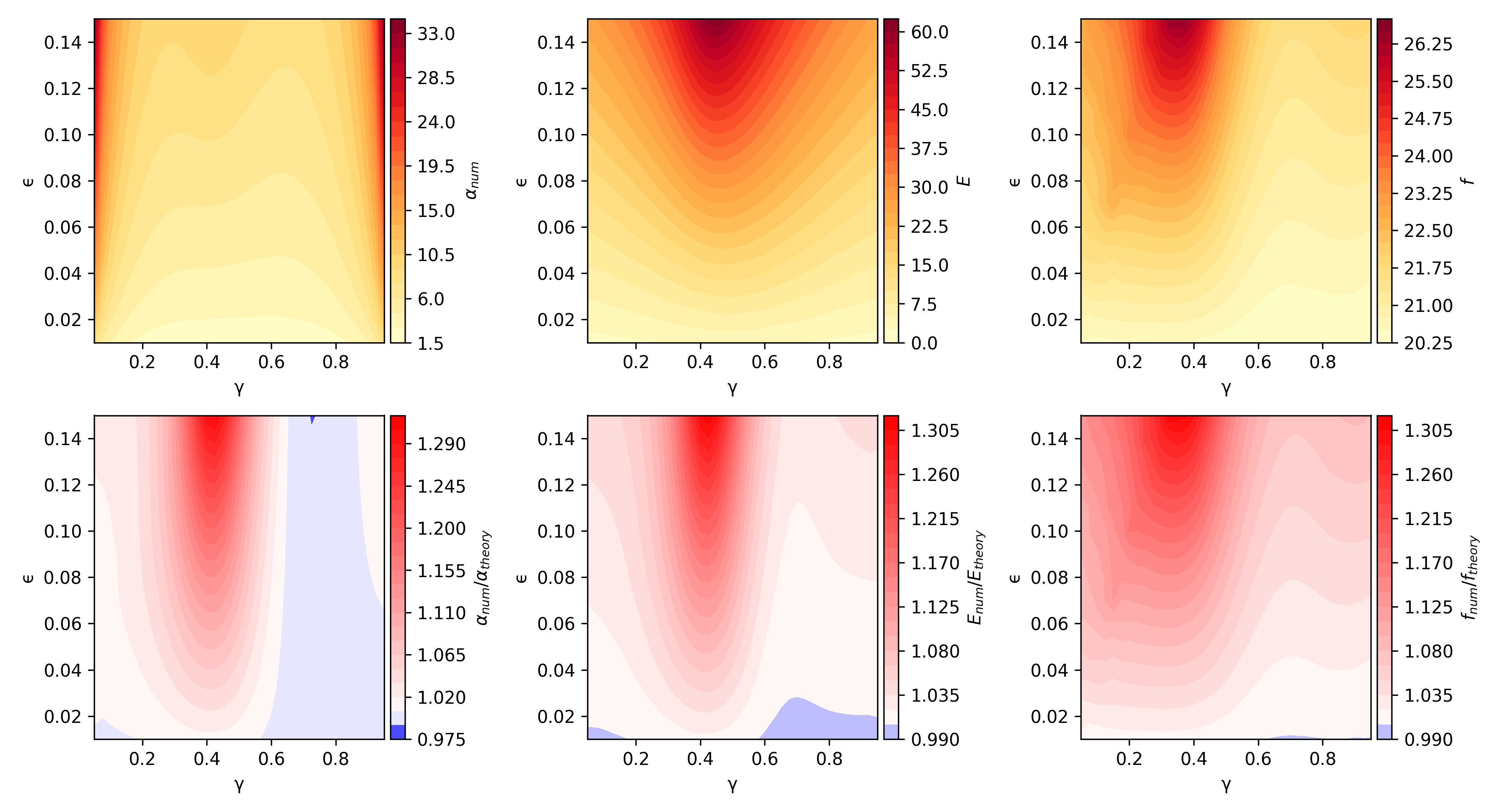}
    \caption{Comparison of snap-through analysis obtained through non-linear solve\_bvp optimization and shallow angle theory, for centered rectangular stimulus profiles. }
    \label{fig: SI non-linear thresholds full}
\end{figure}

Moreover, Fig.\ \ref{fig: SI non-linear thresholds full} shows the agreement of the shallow theory with the fully non-linear approach for a range of $\gamma$ and $\e$, where we can see that the shallow theory gives reasonable results for $\e < 0.05$, with error in the snapping thresholds below $8\%$.

Lastly, the shallow limit introduces additional stimulation profiles that do not affect the beam's behavior.  In the flat-clamping case, the vertical force can offset any linear component in $\kb$, while in the mixed case, $\kb\sim(s-1/2)$ has no influence on the shape as it does not influence the curvature on the pinned end. In the pinned case, the beam remains sensitive to all profiles of preferred curvature.

\subsection{Stimulation with the curvature of pure base mode} \label{app: pure modes}
We denote the compression-normalized solutions to the traditional Euler buckling problem ($\kb=0$) as $\theta_i$,  ($\frac{1}{2}\langle\theta_i^2\rangle=1$). Then, we apply preferred curvature of $\kb = \alpha_i \, \theta_i'$, and consider a following profile function $\theta = \sqrt{\epsilon} \, \theta_i$. Then, the EL equation is
\begin{equation}
    (\sqrt \e \,-\alpha_i) \theta''+f \sqrt \e \, \theta - v = 0.
\end{equation}
which means that for the $\theta_i$ to be the solution to the problem, $f$ must satisfy $f \frac{\sqrt\epsilon}{\sqrt\e-\alpha_i}=f_i$, thus 
\begin{equation}
    \e=\alpha_i^2 \left(\frac{f_i}{f_i-f}\right)^2.
\end{equation}
Therefore, the $i$th mode exists continuously over the whole $f$ range, while at $f=f_j,\, i\neq j$, any amount of $\theta_j$ can also be added to the solution.

\subsection{Linear boundary conditions}
Linear boundary conditions encompass three types of BCs:
\begin{itemize}
    \item flatly clamped --- $\theta\left(\pm\frac{1}{2}\right) =0$
    \item pinned --- $\theta'\left(\pm\frac{1}{2}\right) = 0$
    \item mixed --- $\theta\left(-\frac{1}{2}\right) = 0, \quad \theta'\left(\frac{1}{2}\right) = 0$.
\end{itemize}
In addition to that, they require fixed-end displacement
\begin{gather}
     \langle\theta\rangle=\Delta y = 0,\\
     \frac{1}{2}\langle\theta^2\rangle=\epsilon.
\end{gather}

\subsection{Orthogonality condition for base modes} \label{app: orthogonality}
In the LBC system, we have linearity in all governing equations except for the compression equation. In order for the modes to be independent, we require $\langle\theta\,\varphi\rangle = 0$, if $\theta$ and $\varphi$ are different base modes.

We begin by writing the EL equations for both of the functions and multiplying each EL by the other profile function
\begin{align} 
    \varphi \left(v_{\theta}+\, \theta'' \right) &= - f_\theta \,\theta \, \varphi, \\
    \theta \left(v_{\varphi}+\, \varphi'' \right) &= - f_\varphi \,\theta \, \varphi.
\end{align}
Then, we subtract them to obtain
\begin{equation}
    v_\theta \, \varphi - v_\varphi \,\theta + (\theta'\,\varphi-\theta \, \varphi')'=(f_\varphi-f_\theta) \,\theta \, \varphi,
\end{equation} 
resulting in the condition for mode orthogonality being
\begin{equation}
    (f_\varphi-f_\theta)\langle \theta \, \varphi \rangle = \left(\left((\theta'\,\varphi-\theta \, \varphi')\right|_{-1/2}^{1/2} + v_{\theta} \, \Delta y_\varphi - v_{\varphi} \,\Delta y_\theta) \right),
\end{equation}
with $\Delta y=y(1/2)-y(-1/2)$, thus showing that solutions to LBCs are orthogonal in $\theta$.

Now, we turn our attention to the orthogonality of the curvature. We begin by multiplying the derivatives of EL equations by the compression of the other mode, i.e. 
\begin{align} 
    \varphi' \left(f_\theta \,\theta'+\, \theta''' \right) &= 0, \\
    \theta' \left(f_\varphi  \, \varphi'+\, \varphi''' \right) &= 0.
\end{align}
which we subtract to obtain 
\begin{equation} \label{eq: app curvature orthogonality temp}
    (f_\theta-f_\varphi)\theta'\varphi' = (\theta'\varphi''-\theta''\varphi')',
\end{equation}
in which we substitute $\theta''$ and $\varphi''$ based on EL equations to obtain
\begin{equation}
    (f_\theta-f_\varphi)\langle\theta'\varphi'\rangle = \left.(f_\theta \theta \varphi ' - f_\varphi \theta' \varphi + v_\theta \varphi' - v_\varphi \theta')\right|_{-1/2}^{1/2}.
\end{equation}
While the first two terms can immediately be set to zero for all LBCs, the other two are not immediately apparent. 

For the pinned BCs, we see that the remaining two terms are zero, whereas for the remaining two LBCs, we start by integrating the EL equation
\begin{equation}
    0 = \int_{-1/2}^{1/2}(\theta''+f_\theta \theta + v_\theta) \der s = \theta'(1/2)-\theta'(-1/2)+ f_\theta \Delta y_\theta + v_\theta,
\end{equation}
where $\Delta y_\theta=0$ from BCs, resulting in $\theta'(1/2)=\theta'(-1/2)-v_\theta$.

For the flatly clamped BCs, the even solutions have antisymmetric $\theta(s)$, therefore $\theta'(s)$ is antisymmetric, thus 
\begin{equation}
    \theta'(\pm1/2)= v_\theta=0 
\end{equation}
while the odd solutions have
\begin{equation}
    \theta'(1/2)= -\theta'(-1/2)=-v_\theta / 2 
\end{equation}
In both cases, this results in $\left.(v_\theta \varphi' - v_\varphi  \theta' )\right|_{-1/2}^{1/2} = 0$, proving that flatly clamped modes have orthogonal curvature.

Lastly, for the mixed clamped BCs, we take the $s=1/2$ to be the pinned end which constrains $\theta'(1/2)=0$, resulting in $\theta'(-1/2)=v_\theta$. Therefore, the curvatures of pure modes for mixed BCs are also orthogonal to one another, which proves that Euler modes are orthogonal both in $\theta$ and $\kappa$.

\subsection{Instability at minimum of compression} \label{app: minimum instability}
Perturbing $\theta$, $\lambda$ and $\lambda_v$, by 
\begin{gather}
    \theta(s)\rightarrow\theta(s)+\delta\,  \theta_p(s), \\
    f \rightarrow f + \delta\, f_p, \\
    v \rightarrow v+\delta \,v_{p}.
\end{gather}
gives us the linear stability equation
\begin{equation}
    \theta''_p+f\,\theta_p+ f_p \,\theta - v_{p} =0.
\end{equation}
which is the de facto EL equation differentiated over $f$, i.e.
\begin{equation}
    \frac{\der \theta}{\der f} + f \frac{\der \theta}{\der f} +  \,\theta - \frac{\der v}{\der f} =0. 
\end{equation}
Therefore, $\frac{\der \theta}{\der f}$ solves the EL equation and simultaneously satisfies all boundary conditions at $\frac{\der\e}{\der f}= 0$ as compression constraint becomes $\frac{1}{2}\langle\theta \,\frac{\der \theta}{\der f}\rangle=\frac{\der\e}{\der f}$.

\subsection{Stimulus profiles for the figures in the article}
All profiles presented below are shown exclusive of a multiplicative prefactor $\ak$, peak-to-peak normalized, and correspond to flat clamping.
\begin{align}
    \kb_0= &\frac{1}{2} \cos(2 \pi s / L) ,\\
    \kb_\mathrm{min Thr}= &\mathrm H\left(\frac{s}{L}+0.311\right)-\mathrm H\left(\frac{s}{L}-0.091\right) \\
    &+\mathrm H\left(\frac{s}{L}-0.446\right) ,\\
    \kb_\mathrm{max E}= &\mathrm H\left(\frac{s}{L}+0.158\right) - \mathrm H\left(\frac{s}{L}-0.293\right), \\
    \kb_\mathrm{symm} = &\mathrm H\left(\frac{|s|}{L} - 0.4\right) - \mathrm H\left(\frac{|s|}{L} - 0.2\right),
\end{align}
where $\mathrm H$ is the Heaviside theta.

\begin{table}
    \centering
    \begin{tabular*}{\linewidth}{@{\extracolsep{\fill}}ccccc}
         \hline \hline
         BCs & Profile &  $\ak\, [\sqrt \e]$ & $E_\mathrm{s}\, [\ak^2]$ & $E_\mathrm{s}\, [\e]$ \\ \hline
         \multirow{6}{1.5 cm}{Clamped} 
         &$\kb_0$ & 26.28 & 0.239 & 165.1\\
         &$\kb_1$ & 18.84 & 0 & 0 \\
         &$\kb_\mathrm{rect}|_{\gamma=0.7}$ & 21.52 & 0.279 & 128.9 \\
         % &$\kb_\mathrm{s.critical}$ & 29.49 & 0 & 0  \\
         &$\kb_\mathrm{symm}$ & 22.89 & 0.095 & 49.5 \\  
         &$\kb_\mathrm{max E}$ & 10.33 & 0.661 & 70.6\\
         &$\kb_\mathrm{min Thr}$ & 9.11 & 0.429 & 35.6 \\  \hline
         \multirow{2}{1.5 cm}{Mixed} 
         & $\kb_\mathrm{max E}$ & 8.75 & 0.807 & 61.9 \\
         & $\kb_\mathrm{min Thr}$ & 7.68 & 0.535 & 31.6 \\\hline
         \multirow{2}{1.5 cm}{Pinned} 
         & $\kb_\mathrm{max E}$ & 7.67 & 0.831 & 48.9 \\ 
         &$\kb_\mathrm{min Thr}$ & 6.72 & 0.677 & 30.6 \\ \hline \hline
    \end{tabular*}
    \caption{Snapping energies and thresholds for stimulus profiles presented in the paper.}
    \label{tab: snapping properties}
\end{table}

\subsection{Pure buckling mode profiles } \label{sec: SI buckling modes}
This section contains analytical profiles for all fundamental modes of three types of LBCs, where all modes are compression-normalized and $f_i=k_i^2$. To recover the preferred curvature profile, one can take $\k_i(s)|_{f=0}=\kb(s)$.

\subsubsection{Flatly clamped}
\begin{figure}[h]
    \centering
    \includegraphics[width=\linewidth]{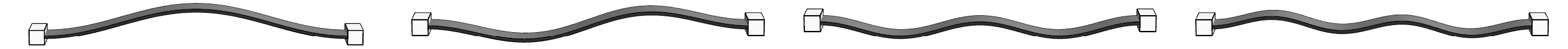}
    \caption{The first four Euler buckling modes for a flat-clamped boundary conditions.}
    \label{fig: flat clamping Euler}
\end{figure}
Even modes have  $\sin(k_i/2)=0$ ($k_{0,2,4,...}=2 \pi, 4\pi, 6\pi,...$) with
\begin{gather}
    y_i(s)=\frac{2 \alpha _i k _i }{f-f_i} \left(\cos \left(\frac{k _i}{2}\right)-\cos \left(k _i s\right)\right), \quad
    \theta_i(s) = \frac{2\alpha _i f_i  }{f-f_i} \sin \left(k _i s\right), \\
    \kappa_i(s)=\frac{2 \alpha _i f_i  k _i }{f-f_i} \cos \left(k _i s\right), \quad
    \e_i = \frac{\alpha _i^2 f_i^2 }{\left(f-f_i\right)^2}, \quad 
    E_i= \frac{\alpha _i^2 f_i f^2  }{\left(f-f_i\right){}^2}
\end{gather}
and the odd modes have $k_i/2 = \tan(k_i/2)$ ($k_{1,3,5,...}=2.86\pi, 4.91\pi,6.94\pi$,...), with 
\begin{gather}
    y_i(s) = \frac{4 \alpha _i }{f-f_i} \left(k _i s-\sec \left(\frac{k _i}{2}\right) \sin \left(k _i s\right)\right), \quad 
    \theta_i(s) = \frac{4 \alpha _i k _i }{f_i-f} \left(\sec \left(\frac{k _i}{2}\right) \cos \left(k _i s\right)-1\right), \\
    \k_i(s)= \frac{4 \alpha _if_i  }{f-f_i} \sec \left(\frac{k _i}{2}\right) \sin \left(k _i s\right), \quad
     \e_i = \frac{\alpha _i^2 f_i^2 }{\left(f-f_i\right)^2}, \quad
     E_i = \frac{\alpha _i^2  f_i f^2 }{\left(f-f_i\right){}^2}.
\end{gather}

\subsubsection{Pinned-pinned}
\begin{figure}[h]
    \centering
    \includegraphics[width=\linewidth]{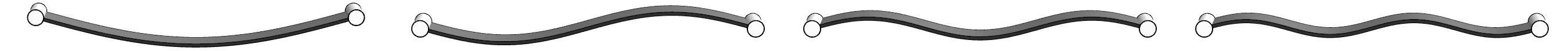}
    \caption{The first four Euler buckling modes for pinned boundary conditions.}
    \label{fig: pinning Euler}
\end{figure}
Even modes have $\cos(k_i/2)=0$ ($k_{0,2,4,...}=\pi,3\pi,5\pi$,...)  with
\begin{gather}
    y_i(s)=\frac{2 \alpha _i k _i }{f-f_i} \cos \left( k _i s\right), \quad
    \theta_i(s)= \frac{2 \alpha _i f_i  }{f-f_i} \sin \left(k _i s\right ), \\
    \kappa_i(s) = \frac{2 \alpha _i f_i  k _i }{f-f_i} \cos \left(k _i s\right), \quad
     \e_i = \frac{\alpha _i^2 f_i^2 }{\left(f-f_i\right)^2}, \quad 
     E_i = \frac{\alpha _i^2 f_if^2  }{\left(f-f_i\right){}^2}.
\end{gather}
Odd modes have $\sin(k_i/2) =0$ ($k_{1,3,5,...}=2\pi,4\pi,6\pi,...$) with
\begin{gather}
    y_i(s)=-\frac{2 \alpha _i k _i }{f-f_i} \sin \left(k _i s\right), \quad 
    \theta_i(s) = -\frac{2\alpha _i  f_i }{f-f_i} \cos \left(k _i s\right), \\
    \kappa_i(s) = \frac{2 \alpha _i f_i k _i }{f-f_i} \sin \left(k _i s\right), \quad 
    \e_i = \frac{\alpha _i^2 f_i^2 }{\left(f-f_i\right)^2} ,\quad
    E_i = \frac{\alpha _i^2 f_i  f^2}{\left(f-f_i\right){}^2}
\end{gather}

\subsubsection{Clamped-pinned}
\begin{figure}[h]
    \centering
    \includegraphics[width=\linewidth]{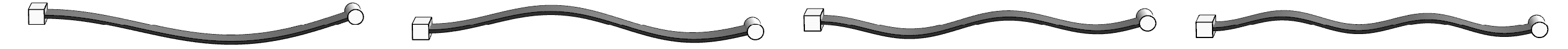}
    \caption{The first four Euler buckling modes for clamped-pinned boundary conditions.}
    \label{fig: clamped pinned Euler}
\end{figure}
Clamped-pinned admit only a single type of solutions with $k_i=\tan(k_i)$ ($k_{0,1,2,3,...}=1.43 \pi,2.46\pi,3.47\pi,4.48\pi$,...) with
\begin{gather}
    y_i(s)=-\frac{2 \alpha _i \sec \left(\frac{k _i}{2}\right) }{f-f_i} \left((2 s-1) \sin \left(k _i\right)+2 \sin \left( k
   _i\left(\frac{1}{2}- s\right) \right)\right), \\
   \theta_i^\mathrm{mix}(s) = \frac{8 \alpha _i \sin \left(\frac{k _i}{2}\right) }{f-f_i} \left(\sec \left(k _i\right) \cos \left( k _i \left(\frac{1}{2}- s\right) \right)-1\right), \\
   \kappa_i(s) = \frac{4  \alpha _i f_i \sec \left(\frac{k _i}{2}\right) }{f-f_i} \sin \left( k _i\left(\frac{1}{2}- s\right) \right),\\
   \e_i = \frac{\alpha _i^2 f_i^2 }{\left(f-f_i\right)^2}, \quad 
   E_i = \frac{\alpha _i^2 f_i  \left(\cos \left(k _i\right)+2\right)  f^2}{\left(f-f_i\right){}^2}.
\end{gather}

\subsection{Numerical Fourier decomposition} \label{sec: SI Fourier}
To find a Fourier decomposition corresponding to a curvature profile $\kb(s)$ defined at equispaced $n$ points, we use $k$ successive buckling modes, which we integrate over the domain by summing the curvature values over $N$ points (obtained through linear interpolation of $\kb$). Typically, we use $n\sim100$, $N=10001$, $k=701$, though only the first few modes capture most of the energy and compression landscapes; however, given the low computational cost, we chose to work with a larger number. Notably, the preferred curvatures that do not result in any deformation are also included to correctly calculate the preferred curvature profile and the elastic energy.

The constructed model provides information on the bistability threshold and enables the calculation of the arch profile for any value of $\ak$ by identifying the corresponding $f$.

\subsection{Solutions for $\kb_\mathrm{rect}$ with flat-clamping} \label{sec: SI rect}
Rectangular stimuli are essential from the perspective of optimal profiles, as they can maximize coupling to Euler buckling modes at a given peak-to-peak amplitude. The equations below describe the response of the stip to central rectangular stimulation of width $\gamma$, where $\lambda=\sqrt{f}$ 

\begin{equation}
\theta(s)=\frac{\ak}{4\lambda^2}
     \begin{cases} 
      -2 \csc (\lambda ) \sin (\gamma  \lambda ) \sin ^2\left(\lambda  \left(s+\frac{1}{2}\right)\right) & s\leq -\gamma/2 \\
      2 \left(\sin ^2\left(\frac{\gamma  \lambda }{2}+\lambda  s\right)-\csc (\lambda ) \sin (\gamma  \lambda ) \sin ^2\left(\lambda  \left(s+\frac{1}{2}\right)\right)\right) & -\gamma/2\leq s\leq \gamma/2 \\
      \csc (\lambda ) \sin (\gamma  \lambda ) (\cos (\lambda -2 \lambda  s)-1) & \gamma/2\leq x 
   \end{cases}
\end{equation}

\begin{equation}
\begin{split}
    \e_\mathrm{rect} =\frac{1}{4} \ak^2 \csc ^2(\lambda ) \sec ^2(\lambda ) (-\sin (4 \gamma  \lambda )-\sin (4 \lambda -4 \gamma  \lambda )-4 \gamma  \lambda  \cos (4 (\gamma -1) \lambda )\\+4 (\gamma -1) \lambda  \cos (4 \gamma  \lambda )+4 \lambda +\sin (4 \lambda ))
\end{split}
\end{equation}

\begin{gather}
\begin{split}
    E=\frac{\ak^2 \csc ^2(\lambda )}{16 \lambda } \big(4 \gamma  \lambda -6 \sin ^2(\lambda ) \sin (2 \gamma  \lambda )-2 \gamma  \lambda  (\cos (2 (\gamma -1) \lambda )+2 \cos (2 \lambda))\\+(2 (\gamma -1) \lambda -3 \sin (2 \lambda )) \cos (2 \gamma  \lambda )+2 \lambda +3 \sin (2 \lambda )\big)
\end{split}
\end{gather}
Alternatively, their associated fundamental mode amplitudes are $\alpha_{i}^{even}=-\frac{\sin(k_i/2)}{k_i^2}$ and $ \alpha_{i}^{odd}=0$.

\begin{figure} [h]
    \centering
    \includegraphics[width=5in]{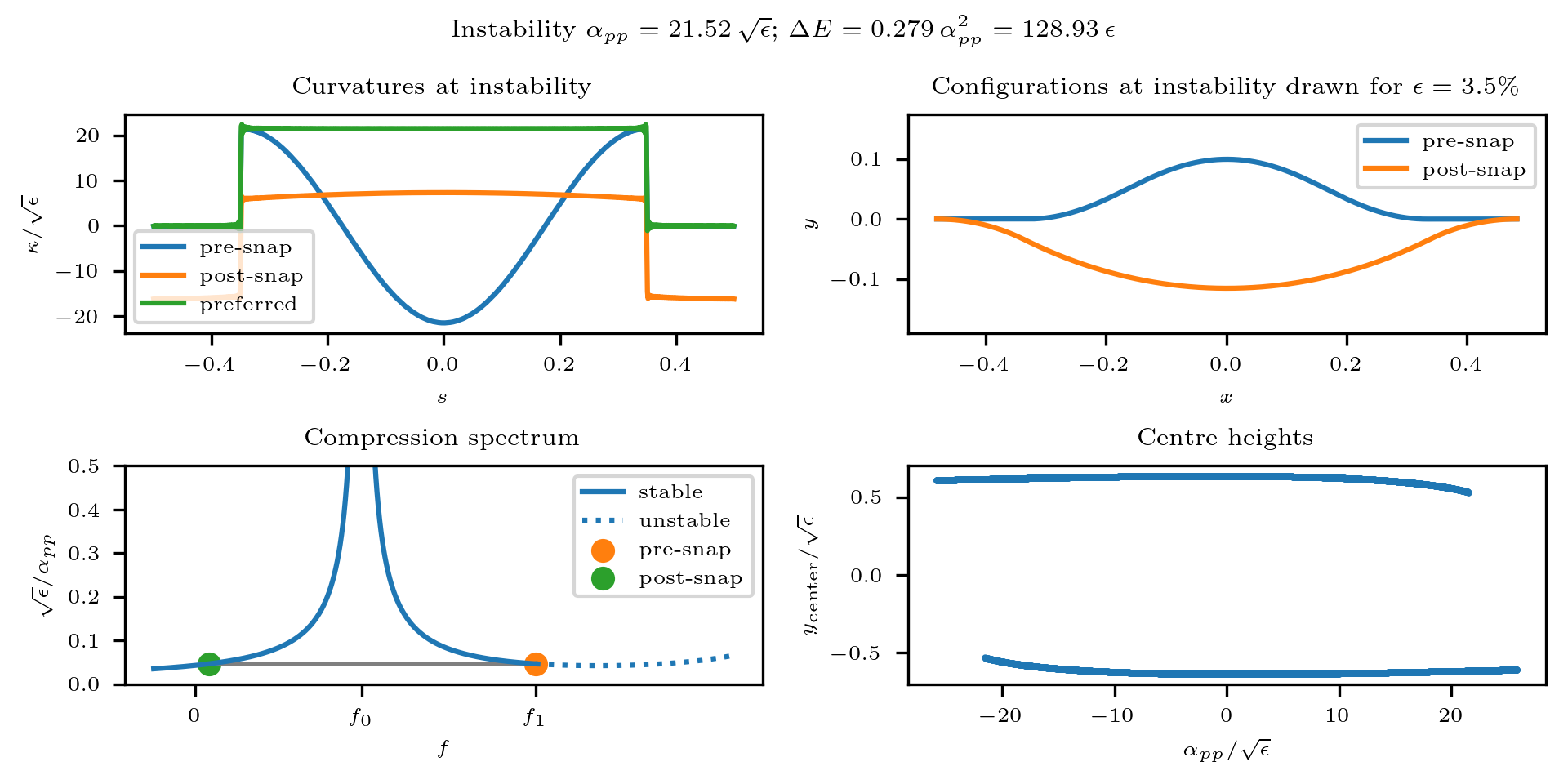}
    \caption{Snapping properties of the centred rectangular preferred curvature with $\gamma=0.7$ obtained through Fourier analysis.}
    \label{fig: SI rect 70}
\end{figure}
Given the compression spectrum, we find that the instability occurs at $f=f_1$, along the asymmetric deformation pathway driven by the first fundamental mode. Fig.\ \ref{fig: SI rect 70} shows the beam pre- and post-snapping profiles together with the corresponding curvatures in addition to the compression spectrum and center heights.  

\begin{figure}[ht]
    \centering
    \includegraphics[width=7in]{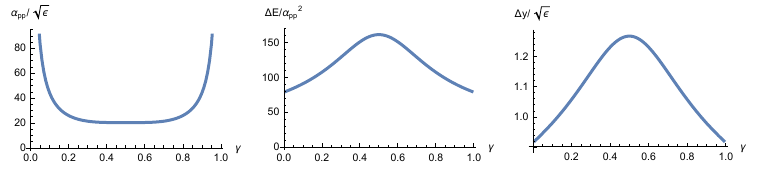}
    \caption{Comparison of snapping threshold, released energy, and center displacements for central rectangular stimulation. }
    \label{fig: SI centred rect}
\end{figure}

For rectangular stimulation, which depends only on a single parameter $\gamma$, we plot the corresponding $\ak(\gamma)$ at the instability threshold and the released energy $\Delta \mathcal E(\gamma)$. We see that the $\gamma=50\%$ is best both in terms of having the lowest $\ak$ and the highest $\Delta E$, as shown in Fig.\ \ref{fig: SI centred rect}. 

However, considering only symmetric stimulation yields an incomplete picture, as it excludes odd modes. At the same time, we know that the presence of the 1st mode can substantially lower the snap-through threshold, and the stimulation couples into the 1st mode through misalignment with respect to the strip's center.

\begin{figure}[ht]
    \centering
    \includegraphics[width=7 in]{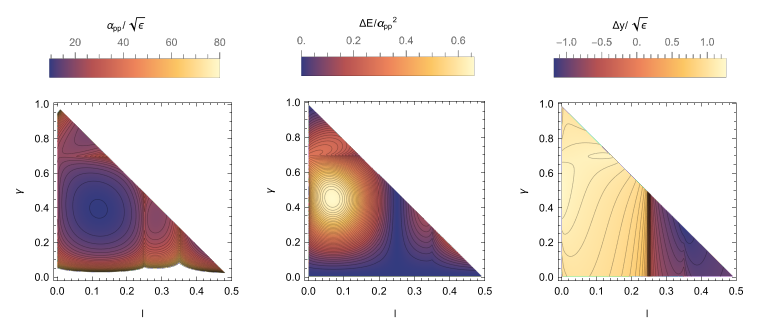}
    \caption{Comparison of snapping threshold, released energy, and center displacements for offset rectangular stimulation. }
    \label{fig: SI offset rect}
\end{figure}
We therefore analyze an offset rectangular stimulation $\kb_\mathrm{rect}(\gamma, l) = \ak \,\mathrm{rect}((s-l)/\gamma)$. Due to the complexity of the formulas, we do not present their analytical form in the text, but the amplitudes of the associated modes are 
\begin{gather}
    \alpha_{i}^{even}= -\frac{\cos(k_i l) \sin(k_i\gamma)}{k_i^2}, \\
    \alpha_{i}^{odd}=-\frac{\sin(k_i l)\sin(k_i \gamma / 2)}{2 k_i \cos(k_i/2)}.
\end{gather}
The plots shown in Fig.\ \ref{fig: SI offset rect} reveal a clear modal structure: at $l = \frac{1}{4}$ the stimulation is centered around a zero of $m_0$, decoupling the stimulation from $m_0$ for any $\gamma$, and no snap-through is possible. This results in the change of snapping direction when $l$ exceeds $\frac{1}{4}$ with a supercritical transition at $l=\frac{1}{4}$, and curvature, energy, and displacement have discontinuous derivatives over this boundary.

Moreover, similarly, the coupling to $m_1$ can be observed. At $l= \frac{\pi}{k_1}\approx0.35$ there is a sharp change in the character of the stimulation and when $\gamma = \frac{2 \pi}{k_1}\approx0.7$, the stimulation decouples from $m_1$ for any value of  $l$, making this stimulus width more experimentally robust, where perfect alignment is difficult to achieve.

\subsection{Continuous transformation around the critical point} \label{sec: SI continuous}
Transition from the ``up'' configuration to the ``down'' configuration does not require a snap-through, as one can traverse the phase diagram from Fig.\ 2 in the mono-stable region, around the critical point back to the origin, effectively returning to the same point on the phase diagram, but in the opposite state. To realize this, one must use a stimulation sequence that continuously transitions from a curvature profile with a positive $\alpha_0$ to one with a negative $\alpha_0$, while remaining away from the critical point at $\alpha_0=0$. 

\begin{figure} [h]
    \centering
    \includegraphics[width=6.4 in]{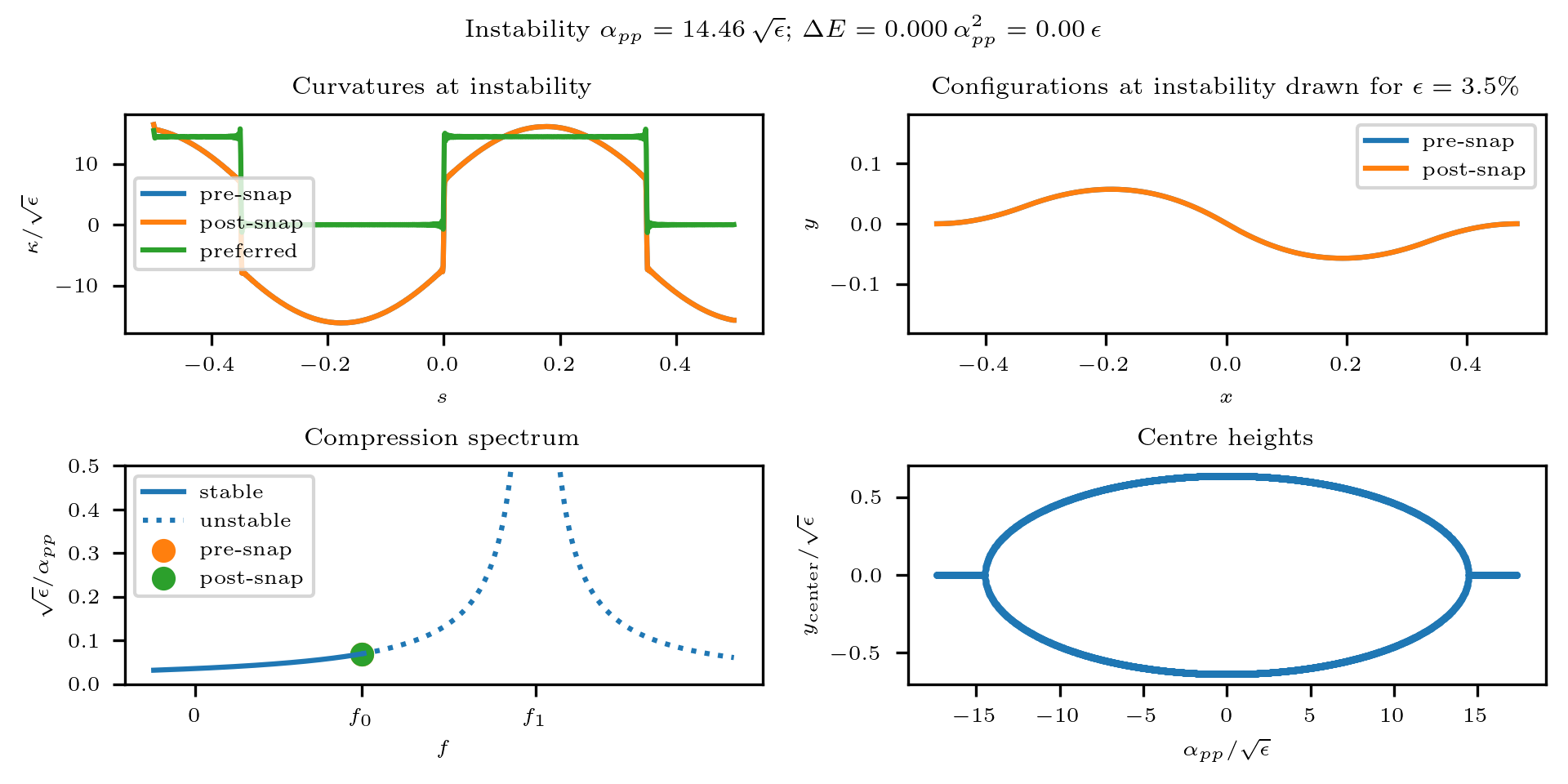}
    \caption{Instability properties of the binarized $m_1$ mode. }
    \label{fig: SI binarised first mode}
\end{figure}
We chose to work with a stimulus that continuously transitions from a binarized $m_0$ to a binarized $m_1$, as shown in Mov.\ 3. These stimuli maximize coupling to the $m_0$ and $m_1$ at any given value $\ak$, which moves the critical point to the lowest possible $\ak=14.46\sqrt \e$, Fig.\ \ref{fig: SI binarised first mode}. We targeted the lowest threshold for the supercritical transition, as the further we are from the critical point, the smoother the deformation will be. 

\begin{figure} [ht]
    \centering
    \includegraphics[width=4.8in]{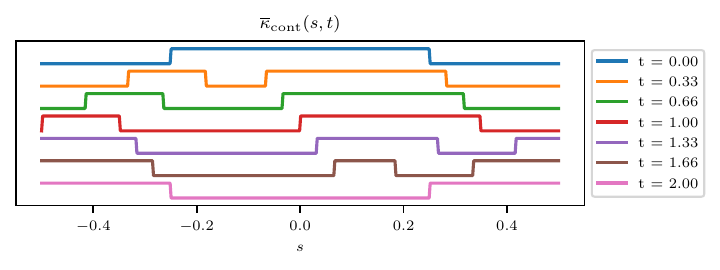}
    \caption{Snapshots of the stimulus for continuous transition around the critical point at different values of control parameter $t$.}
    \label{fig: SI continuous profiles}
\end{figure}

The exact formula for the deformation is 
\begin{equation}
    \kb_\mathrm{cont}(s,t < 1)= \ak\begin{cases}
        1 & \text{for }-\frac{1}{4}(1 + t) < s< -\frac{1}{4} (1 - t)+w_2\\
        1 & \text{for }s_0 (1-t)<s< s_0 (1-t) + w_1,   \\
        0 & \text{otherwise}, 
    \end{cases}
\end{equation}
where 
\begin{gather}
    w_1=\pi /k_1, \quad 
    w_2=\frac{1}{2}-w_1, \quad
    s_1=\frac{1}{4}-w_1, \\
    \kb_\mathrm{cont}(s, t > 1) = 1 - \kb_\mathrm{cont}(-s, 2-t).
\end{gather}
The snapshots of the stimulation at multiple points in time are shown in Fig.\ \ref{fig: SI continuous profiles}.

\begin{figure} [h]
    \centering
    \includegraphics[width=4.8in]{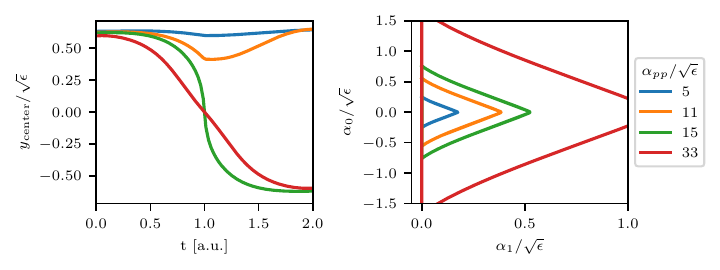}
    \caption{Continuous transition for the $\kb_\mathrm{cont}$ for different values of $\ak$, both below and above the critical point. }
    \label{fig: SI critical point loop}
\end{figure}

\begin{figure}
    \centering
    \includegraphics[width=0.5\linewidth]{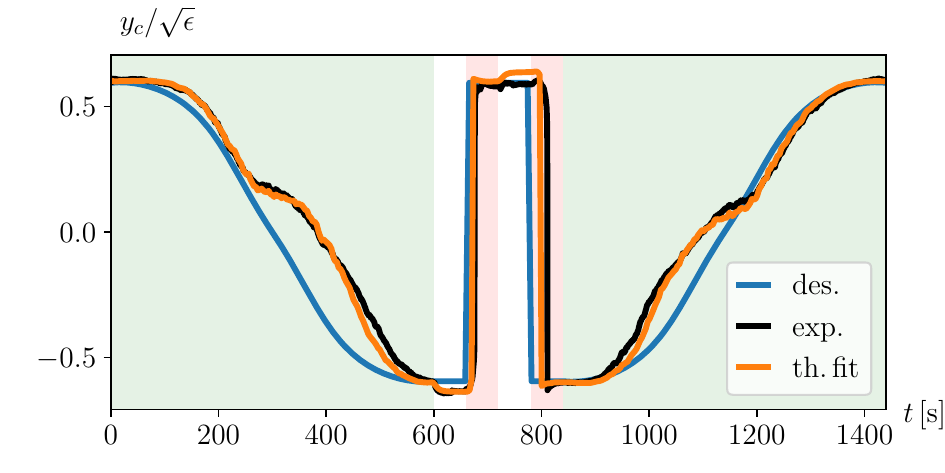}
    \caption{Center height of the beam stimulated with the continuous transition profile. }
    \label{fig: SI continuous centre height}
\end{figure}

 Experimentally, at $\e\approx1.2\%$ the maximum achievable strength is $\ak\approx33\sqrt \e$, which is far enough from the threshold (critical point) that the center-displacement will be smooth. The comparison of the center displacements for various values of attainable $\ak$ is shown in Fig.\ \ref{fig: SI critical point loop}. The center-height-time plot is shown in Fig.\ \ref{fig: SI continuous centre height}, where we compare the theoretical design with the direct and indirect experimental measurements. The direct measurements use optical data, whereas the indirect measurements use a Fourier decomposition of the extracted curvature profile to predict the center height, which agrees well with the direct measurements.

\subsection{Optimization procedure and results} \label{sec: SI optimisation}
The optimization occurs in two stages --- numerical and analytical. The purpose of the numerical optimization is to identify the optimal curvature profile, first at a coarse stage and then at a fine stage, which is then refined analytically.

The numerical optimization starts from a coarse global optimization using a curvature profile defined at $n=26$ and using \emph{dual annealing} routine from SciPy optimize package \cite{Scipy_2020}, which is then followed by a coarse local search using the \emph{minimize} routine from the same package, either using ``Powell'' or ``Nelder-Mead'' methods. Then, the same routine is used for fine local optimization, in which the number of nodes along which $\kb(s)$ is defined is increased to $n=101$, and the coarse solution is used as a starting point. 

The result is then rewritten as an analytical expression in Mathematica \cite{Mathematica_2024}, and further local optimization is conducted via the \emph{NMinimize} routine, with the initial guess obtained from numerical analysis. 

\begin{figure}[h]
    \centering
    \includegraphics[width=6.4in]{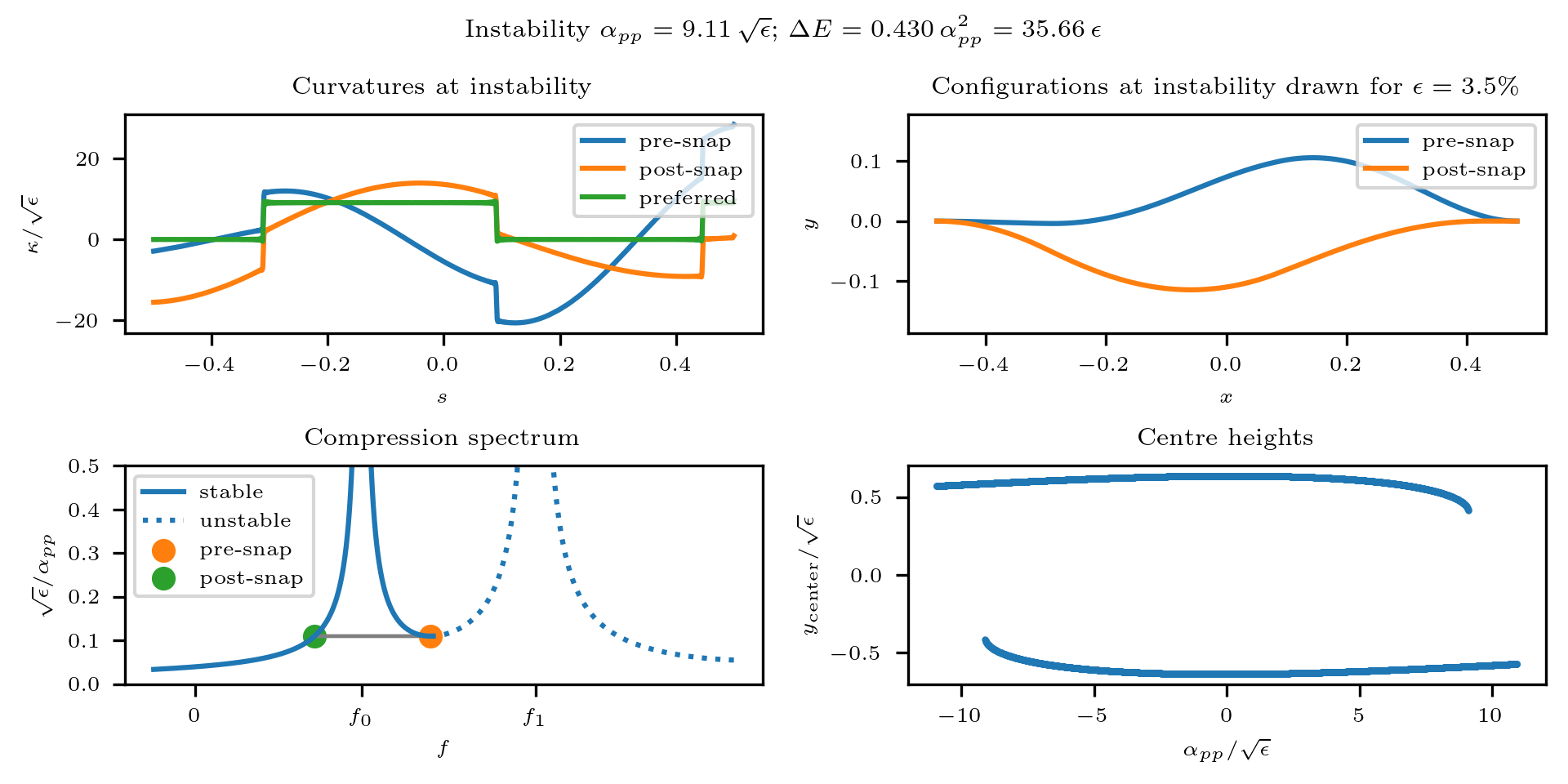}
    \caption{Snapping properties of the curvature minimized stimulus for flat-clamped BCs.}
    \label{fig: SI minimum curvature}
\end{figure}

\begin{figure}[h]
    \centering
    \includegraphics[width=6.4in]{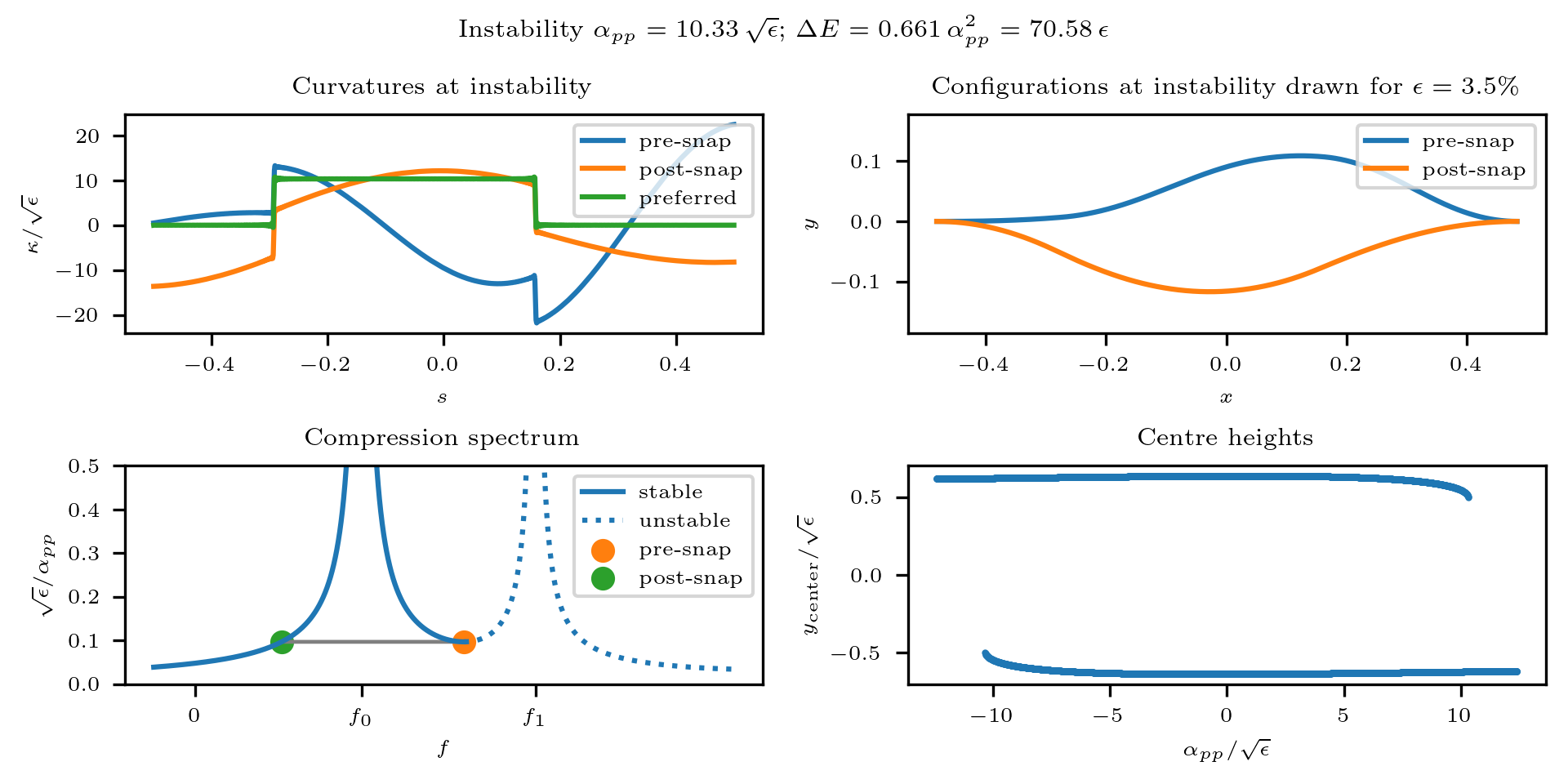}
    \caption{Snapping properties of the energy maximizing stimulus for flat-clamped BCs.}
    \label{fig: SI maximum curvature}
\end{figure}

\subsection{Applying the model to passive systems}
While previous sections focused on stimulus control ($\kb \propto\alpha$) at fixed compression ($\epsilon$), the actual control parameter governing snapping is $\sqrt{\epsilon}/\alpha$, so exactly the same set of behaviors can be observed  by controlling the arch-compression $\epsilon$ at fixed $\alpha$. Our framework thus also directly applies to passive arches fabricated with a fixed preferred curvature ($\kb =\alpha f(s) L^{-1}$) and then switched by changing $\epsilon$
--- i.e.\ by moving the positions of the clamped ends.

Such passive arches can be created following any of the curvature patterns analyzed previously, and then analyzed using the same ``compression spectrum'' graphs ($\sqrt{\epsilon}/\alpha$ vs horizontal force $f$) previously presented. Identically to the $\alpha$ controlled case, the threshold compression delineating mono- and bi-stability for such arches (and hence the snapping threshold)is then found either at $f_1$ if $\frac{\partial \epsilon}{\partial f}|_{f_1} < 0$, or at compression-minimizing force $f_m$ between $f_0$ and $f_1$ otherwise. However, in the passive case, it is natural to write the threshold compressions as a fraction of the fabrication compression $\epsilon_{fab} =\epsilon|_{f=0}\propto \alpha^2$, which is the natural (horizontally free) compression of the arch, and corresponds to the intersection of the compression spectrum with the $y$ axis. A list of such thresholds for the key cases considered in the paper is given in Tab.\ \ref{tab: passive thresholds}.

\begin{table}[h]
    \centering
    \begin{tabular}{c|c}
        Mode & $\epsilon_{th} / \epsilon_{fab}$ \\ \hline
        fundamental (0th) mode & 0.914 \\
        1st mode & 3.822 \\
        binary central 70\% & 1.176 \\
        curvature minimizing & 7.691 \\
        energy maximizing & 4.095 \\
        symmetric transition & 2.387 
    \end{tabular}
    \caption{Threshold compressions for mono-/bi-stability in passive analogues of curvature profiles used earlier in the manuscript. }
    \label{tab: passive thresholds}
\end{table}

Notably, out of all the curvature profiles considered, only the pure fundamental mode is bistable at fabrication compressions (Tab.\ \ref{tab: passive thresholds}, $\epsilon_{th} / \epsilon_{fab} < 1$), while all the remaining profiles must be compressed more to obtain bistability, and then snap in subsequent decompression. We also highlight that not only the thresholds, but also the modes of instability are the same as in the curvature-stimulated case, so snap-through occurs if the beam's preferred curvature contains a $\kappa_0$ component (which can be identified by a nonzero center height during fabrication) while otherwise a supercritical transition is observed. Thus, much of the richness of curvature stimulation carries over to passive by naturally curved beams, although, of course, a naturally curved beam is limited to only one preferred curvature profile, so a different beam is needed for each behavior.

\bibliographystyle{abbrv}
\bibliography{references}